\documentclass[prd, aps, superscriptaddress, preprintnumbers, twocolumn, floatfix, nofootinbib]{revtex4}

\pdfoutput=1

\usepackage{amsfonts}

\usepackage{amsmath}

\usepackage{amssymb}

\usepackage{bm}

\usepackage{dcolumn}

\usepackage{graphicx}

\usepackage[latin1]{inputenc}

\usepackage{latexsym}

\usepackage{rotating}

\usepackage{hyperref}

\usepackage{graphicx}

\usepackage{color}

\newcommand\be{\begin{equation}}

\newcommand\ba{\begin{eqnarray}}

\newcommand\ee{\end{equation}}

\newcommand\ea{\end{eqnarray}}

\newcommand{\Msol}{\ensuremath{M_{\odot}}}

\begin{document}

\title {Intermediate Mass Black Hole Seeds from Cosmic String Loops}

\author{Robert Brandenberger}

\email{rhb@physics.mcgill.ca}

\affiliation{Department of Physics, McGill University, Montr\'{e}al, QC, H3A 2T8, Canada}

\author{Bryce Cyr}

\email{bryce.cyr@mail.mcgill.ca}

\affiliation{Department of Physics, McGill University, Montr\'{e}al, QC, H3A 2T8, Canada}

\author{Hao Jiao}

\email{hao.jiao@mail.mcgill.ca}

\affiliation{Department of Physics, McGill University, Montr\'{e}al, QC, H3A 2T8, Canada}

\date{\today}


\begin{abstract}

We demonstrate that cosmic string loops may provide a joint resolution of two mysteries surrounding recently observed black holes. For a string tension in an appropriate range, large radius string loops have the potential to provide the nonlinearities in the early universe which seed supermassive black holes. The more numerous smaller radius string loops can then seed intermediate mass black holes, including those with a mass in the region between $65 \Msol$ and $135 \Msol$ in which standard black hole formation scenarios predict no black holes are able to form, but which have recently been detected by the LIGO/VIRGO collaboration. We find that there could be as many as $10^6$ of intermediate mass black holes per galaxy, providing a tantalizing target for gravitational wave observatories to look for.

\end{abstract}


\pacs{98.80.Cq}

\maketitle


\section{Introduction} 

\label{sec:intro}

In this Letter we suggest that cosmic strings may provide a joint resolution of two puzzles in astrophysics. On one hand, as already investigated in \cite{Jerome}, string loops may explain the origin of the seeds about which super-massive black holes (SMBHs) accrete. On the other hand, smaller loops can lead to the formation of intermediate mass black holes, in particular black holes  in the ``mass gap'' range between $65 \Msol$ and $135 \Msol$ where standard stellar black hole formation scenarios predict that no black holes should exist. Such black holes, however, have been detected by the LIGO/VIRGO collaboration \cite{LIGO}.

Observations indicate the presence of black holes of mass larger than $10^9 \Msol$ at redshifts greater or equal to $z = 6$ \cite{obs}. In fact, each galaxy in the low redshift universe appears to harbour a SMBH of mass greater or equal to $10^6 \Msol$. In the context of the current cosmological paradigm, the $\Lambda$CDM model with an almost scale-invariant spectrum of nearly Gaussian primordial density fluctuations, it is not possible to explain the origin of these massive early black holes if accretion is limited by the Eddington rate \cite{Jerome}. For a review article on supermassive black hole formation the reader is referred to \cite{Marta}. As studied in \cite{Jerome}, cosmic string loops with an appropriate radius can provide nonlinear seeds at high redshifts which can resolve this puzzle.

Cosmic strings exist as solutions of the field equations in a wide range to particle physics models beyond the {\it Standard Model}. If Nature is described by such a theory, then a network of strings inevitably forms in the early universe and persists to the present time. For a review article the reader is referred to \cite{CSrevs}. The network of strings contains loops with a continuous range of radii. String loops represent nonlinear density fluctuations. Hence, string loops may seed black holes with a continuous range of masses. As will be reviewed in the next section, the number density of string loops increases as the loop radius decreases. If the parameters of the string model are tuned such that the observed number density of super-massive black holes results, the model will predict a distribution of black holes of intermediate mass, in particular of mass in the ``mass gap'' region. Here, we compute the mass distribution of the resulting black hole seeds and show that they may explain the LIGO/VIRGO data.

In the following section we briefly review how cosmic strings form and evolve in an expanding universe. In Section 3 we discuss the mystery of the origin of SMBHs and the possible role which Eddington accretion about string loops can play. Our analysis is summarized in Section 4. The string network is determined by a single free parameter, the string tension. We fix this parameter such that we obtain the observed number density of SMBHs. We then compute the predicted distribution of nonlinear seeds of smaller mass which may evolve into intermediate mass black holes (IMBHs). We conclude with a summary and discussion of our results.

We will use natural units in which the speed of light, Planck's constant and Boltzmann's constant are all set to $1$. We work in the context of a homogeneous and isotropic background metric with scale factor $a(t)$ ($t$ being time). The present time is denoted by $t_0$, the time of equal matter and radiation by $t_{eq}$ (and the corresponding temperatures are $T_0$ and $T_{eq}$, respectively). Instead of time, we will often use cosmological redshift $z(t)$ given by

\begin{align}
z(t) + 1 \, \equiv \, \frac{a(t_0)}{a(t)} \, .
\end{align}

Newton's  gravitational constant is denoted by $G$, and it defines the Planck mass $m_{pl}$ via $G = m_{pl}^{-2}$. The Hubble radius is the inverse expansion rate and plays a role in the description of the network of strings.

\section{Cosmic String Formation and Evolution} \label{review}

A subset of particle physics models beyond the Standard Model admit solutions of their field equations which correspond to linear topological defects analogous to vortex lines in superconductors and superfluids \cite{CSrevs}. If Nature is described by such a model, then a network of strings inevitably \cite{Kibble} forms in the early universe and persists to the present time. Typically, topological defects form during a symmetry breaking phase transition when a scalar order parameter takes on a nonvanishing expectation value. If the manifold of possible low temperature expectation values of the order parameter has the topology of a circle, then the defects are one-dimensional strings. They represent narrow tubes of trapped energy \footnote{In this paper we will assume that the strings are not superconducting.}, and they are characterized by their tension $\mu$. The tension is related to the energy scale $\eta$ of symmetry breaking via $\mu \sim \eta^2$. The trapped energy density leads to gravitational effects which in turn produce distinctive signals for strings in various observational windows (see e.g. \cite{RHBCSrev}).

Cosmic strings cannot have any ends. The network of strings which forms in the symmetry breaking phase transition consists of {\it long} strings (strings whose curvature radius is larger than the Hubble radius) and loops. The network rapidly approaches a {\it scaling solution} in which the statistical properties of the strings are the same at all times if all length are scaled to the Hubble radius. The scaling solution is maintained by the long strings intersecting and giving off energy in the form of string loops. The string loops, in turn, oscillate, emit gravitational radiation and gradually decay. There are good analytical arguments to expect the string network to scale (see e.g. \cite{CSrevs}). The scaling of the network of long strings is also clearly established based on numerical simulations \cite{LSsimuls}. Simulations making use of the Nambu-Goto effective action for strings also establish the scaling of the loop distribution \cite{NGsimuls}, while some simulations of cosmic string evolution using the field theory equations \cite{Hind} indicate that the long strings more efficiently lose energy to particles, and that in consequence the distribution of string loops does not scale. We will here assume that the results of the Nambu-Goto simulations are correct.

According to the one-scale model \cite{onescale} (supported by the Nambu-Goto simulations), the number density per unit radius of loops at any given time $t > t_{eq}$ is given by
\begin{align} 
n(R, t) \, = \begin{cases}
 \, N R^{-5/2} t^{-2} t_{eq}^{1/2} \,\,\, \hspace{2mm} R \geq \gamma G\mu t \label{stringdist} \\  {\rm{const}} \,\,\, \hspace{17.5mm} R < R_{gw} \equiv \gamma G\mu t \\
\end{cases}
\end{align}
where $N$ and $\gamma$ are constants. Here, $R$ is the radius of the loop and $n(R, t)$ gives the number density of loops per $R$ interval \footnote{Note that multiplying $n(R, t)$ by $R$ yields the number density of loops with radius greater or equal to $R$, which is dominated by loops of radius between $R$ and $2R$.}. $R_{gw}$ is the radius below which a loop will live less than one Hubble expansion time before decaying. The constant $N$ is determined by the number of long strings per Hubble volume and by the length of loop (in units of the Hubble radius) when it forms, while the constant $\gamma$ is determined by the strength of gravitational radiation from string loops. Based on the results of numerical simulations \cite{NGsimuls} we will take $N \sim 6 \times 10^{-3}$ and $\gamma \sim 10^2$. Note that the above formula is modified for loops which form after $t_{eq}$ and reads 
\begin{align}
n(R, t) \, \sim \,  N' R^{-2} t^{-2} \, . \label{matterdist}
\end{align}
where $N'$ is of the same order of magnitude of $N$. These loops, however, will only play a role for the final considerations in this work.

String loops can also lose energy by {\it cusp annihilation} \cite{RHBcusp}. As can be shown \cite{KT}, for any loop of radius $R$ there will be at least one cusp formed per oscillation time $R$. A cusp is a point on the string - treated according to the Nambu-Goto action - which moves at exactly the speed of light. At this point, the loop doubles back on itself. Since strings have a finite thickness $w \sim \eta^{-1}$, there will be a region near the cusp where the string segments on either side of the cusp point overlap. Locally, this region looks like a string-antistring configuration, and it will hence decay explosively giving rise to a burst of particles. The overlap region has length $l_c \sim w^{1/2} R^{1/2}$ \cite{Olum}, and hence the energy lost by cusp annihilation per unit time is of the order 
\begin{align}
P_{\rm{cusp}} \, \sim w^{1/2} \mu R^{-1/2} \, .
\end{align}
In comparison, the energy loss per unit time of a string loop due to gravitational radiation is \cite{VV}
\begin{align}
P_{\rm{grav}} \, \sim \, \gamma G \mu^2 \, .
\end{align}
Hence, for small values of $\mu$, cusp annihilation will dominate. The critical value $\mu_c$ of $\mu$ below which cusp annihilation dominates depends on the string radius $R$ and is
\begin{align}
G \mu_c \, \sim \, \gamma^{-4/5} G^{1/5} R^{-2/5} \, .
\end{align}
Thus, the larger the loop radius, the less is the relative importance of cusp evaporation. For loops at the gravitational radiation cutoff $R_{gw}$, the condition on $\mu_c$ yields
\begin{align}
G \mu_c \, \sim \, \gamma^{-6/7} \left(\frac{T_{eq}}{m_{pl}} \right)^{4/7} \, \sim \, 10^{-17} \, ,
\end{align}
where we have used the value $\gamma = 10^2$. For values of $G \mu$ smaller than this critical value, the loop distribution changes compared to (\ref{stringdist}). However, in this work we will not be considering values of $G \mu$ smaller than $G \mu_c$.

Note that string loops are not exactly circular. We will introduce a constant $\beta$ to parametrize the mean length $l(R)$ of a loop of radius $R$, namely
\begin{align}
l(R) \, \equiv \, \beta R \, .
\end{align}
For circular loops, $\beta = 2\pi$, though in general one expects $\beta \sim \mathcal{O}(10)$. Since cosmic strings carry energy, their gravitational effects lead to imprints in many observational windows. These imprints are highly non-Gaussian and typically most visible in position space maps. Well known are the line discontinuities which long strings produce in cosmic microwave background (CMB) temperature maps \cite{KS}. The current upper bound on the string tension from not having observed these signals is \cite{CSbound} $G \mu < 10^{-7}$, and searches for these signals using wavelet statistics \cite{Hergt} and machine learning methods \cite{Oscar} have the potential of reducing this bound by one or two orders of magnitude. Long strings moving through space produce {\it wakes}, planar density perturbations in the plane mapped out by the moving string \cite{wake}. Wakes, in turn, lead to specific signals in B-mode CMB polarization maps: rectangles in the sky with a uniform polarization direction and linearly increasing signal amplitude \cite{Holder1}. They also lead to wedge-shaped regions of extra absorption in 21cm redshift maps during the dark ages \cite{Holder2}.

String loops lead to spherical (if the center of mass velocity is small) or filamentary (if the center of mass velocity is large) overdensities. Originally, this mechanism was postulated to be the dominant source of structure formation \cite{original}, but the required value of $G \mu$ exceeds the abovementioned upper bound. Thus, string loops are only a supplementary source of nonlinear structures. The role of string loops in seeding ultra-compact mini-halos was explored in \cite{Maddy}, and the role in seeding globular clusters was studied in \cite{Lin}. Here, we study the role of string loops in seeding SMBHs and IMBHs.

The tightest current constraints on the string tension come from pulsar timing limits on the amplitude of stochastic gravitational waves \cite{MPA}. The limits come about since string loops decay by emitting gravitational radiation resulting in a scale-invariant spectrum of gravitational waves over a large range of wavelengths (with specific signatures coming from the cusp annihilation process). The current limits are $G \mu < 10^{-10}$, and the recent NANOgrav results could \cite{Olum2} be interpreted as being due to cosmic strings with a value of $G\mu$ of this order of magnitude.  The cusp annihilation process also produces jets of particles whose role in contributing to the spectrum of high energy cosmic rays was explored in \cite{Jane}, and which will contribute to the global 21cm signal \cite{CS21cm}, which may play a role in explaining Fast Radio Bursts \cite{Bryce}, and in magnetogenesis \cite{Xinmin}.

Since cosmic strings inevitably arise in a large class of particle physics models beyond the Standard Model, searching for string signals in new observational windows is an interesting way to probe particle physics. Since many of the string signals grow in amplitude as the energy scale $\eta$ increases, cosmology provides an approach to probe particle physics models which is complementary to accelerator probes (which are more sensitive if the energy scale $\eta$ of the new physics is low). Improved upper bounds on the string tension from cosmology will allow us to constrain larger sets of particle physics models (for more discussion on this point see \cite{RHBCSrev2}.

\section{Eddington Accretion and Super-Massive Black Hole Formation} \label{analysis}

The origin of SMBHs is an important open question in astrophysics. A conservative approach (see e.g. \cite{Marta} for a review) is to assume that the seeds of the SMBHs are black holes formed after the death of Population III stars which are expected to have masses in the range of $10^2 - 10^3 \Msol$. These seed black holes are then assumed to accrete matter. The Eddington rate is often taken to be a good estimate for the highest accretion rate onto these seed black holes. But, according to the canonical $\Lambda$CDM paradigm of early universe cosmology, there are not enough nonlinear seeds at early times in order to explain the origin of the observed $10^9 \Msol$ black holes at redshifts greater than $z = 6$. As pointed out in \cite{Jerome}, string loops can provide a sufficient number of nonlinear seeds in the early universe, even for small values of the string tension. For values of the string tension {color{red} significantly lower than the} current upper bounds, linear accretion onto the string loops is insufficient to explain the high mass of the observed SMBHs, and thus it is reasonable to assume that nonlinear accretion at a rate comparable to the Eddington rate takes place. We will denote by ${\cal{E}}_1$ the enhancement of the increase in mass for these SMBH seeds compared to linear theory.

Since there is a continuous distribution of string loop masses with a number density which increases as the mass decreases, the string model for seeding SMBHs predicts a distribution of black holes of smaller masses, and in particular black holes in the ``mass gap'' mass range. Since accretion onto smaller loops is more difficult than accretion onto larger loops, we expect the nonlinear accretion factor for smaller loops to be less. The reason is that for smaller black hole masses, the horizon area is less, and infalling matter has to be directed more precisely in direction towards the black hole, while the thermal correlation length of the accreting matter is independent of the loop size. We will denote that enhancement factor for IMBHs by ${\cal{E}}_2$, noting that it is reasonable to expect that ${\cal{E}}_2 \sim 1$, indicating a scenario in which no Eddington accretion has taken place onto these smaller black holes.

\section{Resulting Distribution of Intermediate Mass Black Holes}

In this section we will compute the expected number density of string-seeded IMBH assuming that the string model is normalized such that it yields one SMBH seed per galaxy. More specifically, we will demand that the model yield one string loop per volume $d_g^3$ (where $d_g$ is the comoving radius of the region which collapses to form a large galaxy) capable of seeding a SMBH of mass $M_{SM}$ at $z=0$, which we will take to be $10^6 \Msol$ later in the analysis. When inserting numbers we will use $d_g = 10^{2/3} {\rm{Mpc}}$ \cite{Marta2}. Allowing for an enhancement of the accretion onto the string loop by a factor of ${\cal{E}}_1$ (e.g. by Eddington accretion) compared to the linear perturbation theory growth rate), the condition on $G \mu$ becomes
\begin{align}
\beta \mu R (z_{eq}+1) {\cal{E}}_1 \, = \, M_{SM} \, ,  \label{mass}
\end{align}
where the radius $R$ must be chosen in order to obtain the correct number density of loops, i.e. (see (\ref{stringdist}))
\begin{align}
N R^{-3/2} t_{eq}^{1/2} t_0^{-2} d_g^3 \, = \, 1 \, . \label{number}
\end{align}

Combining (\ref{mass}) and (\ref{number}) yields 
\begin{align}
G \mu \, = \, \beta^{-1} N^{-2/3} (z_{eq}+1)^{-1/2} {\cal{E}}_1^{-1} \left( \frac{t_0}{d_g} \right)^2 \frac{G}{ t_0}  M_{SM}\, .
\end{align}
Inserting the values of $G$, $d_g$,  $M_{SM}$, and $t_0$, and using $\beta = 10$ yields
\begin{align}
G \mu \, \sim \, 2 \times 10^{-14} N^{-2/3} {\cal{E}}_1^{-1} \, . \label{muvalue}
\end{align}

The mass gap we are interested in today is $[M_{IM}^{\rm{min}}, M_{IM}^{\rm{max}}]$ with $M_{IM}^{\rm{min}} = 65 \, M_{\odot}$ and $M_{IM}^{\rm{max}} = 130 \, M_{\odot}$. As one would expect, black holes in this mass range are also seeded by string loops that fall within a range, $R_{IM}^{min}$ and $R_{IM}^{max}$. Assuming all black holes in this mass range have the same Eddington factor, $\mathcal{E}_2$, (where $\mathcal{E}_2 \leq \mathcal{E}_1$), their masses grow as
\begin{align}
M_{IM}^{\rm{min/max}} \, = \, \beta \mu (z_{eq}+1) R^{\rm{min/max}}_{IM} {\cal{E}}_2 \, , \label{mass2}
\end{align}
The resulting number $N_{IMBH}$ of string loops inside a galaxy capable of seeding  black holes in the IMBH range is then given by
\begin{align}
N_{IMBH} \, \sim \, N  t_{eq}^{1/2} t_0^{-2} d_g^3 R^{\rm{min} \, -3/2}_{IM} \left( 1- \left(\frac{R_{IM}^{\rm{min}}}{R_{IM}^{\rm{max}}} \right)^{3/2} \right)\, , \label{number2}
\end{align}
In fact, it is reasonable to use ${\cal{E}}_2 = 1$. Solving (\ref{mass2}) for $R^{\rm{min/max}}_{IM}$ and inserting the value of $G \mu$ from (\ref{muvalue}) then gives our main result 
\begin{align}
N_{IMBH} \, \sim \, \left(\frac{\mathcal{E}_2}{\mathcal{E}_1}\right)^{3/2} \left(\frac{M_{SM}}{M_{IM}^{\rm{min}}} \right)^{3/2} \left( 1- \left(\frac{M_{IM}^{\rm{min}}}{M_{IM}^{\rm{max}}} \right)^{3/2} \right)\,  \label{IMBHnumber}
\end{align}
We can see that requiring one SMBH per galaxy completely sets the shape of the number distribution of all other black holes formed from cosmic strings, as we would expect in the one-scale model (aside from the different Eddington accretion factors). The number of IMBHs per galaxy is mostly determined by the ratio of SMBH to IMBH masses, with a small correction coming from the small but finite range of IMBHs in the mass gap. Inserting our numbers yields our main prediction of IMBHs per galaxy
\begin{align}
N_{IMBH} \sim 10^6 \left(\frac{\mathcal{E}_2}{\mathcal{E}_1}\right)^{3/2}
\end{align}

If we assume that no loop accretes matter at a rate larger than what linear theory predicts, then we need a value of $G \mu \sim 10^{-13}$ in order to explain the origin of the SMBHs (taking $N^{-2/3} = 10$), and we obtain $N_{IMBH} \sim 10^{6}$. This value of $G \mu$ is consistent with the upper bound on $G \mu$ due to gravitational radiation constraints \cite{MPA}. No super-linear (in particular no Eddington) accretion is required to explain the origin of the nonlinear seeds required to explain the abundance of SMBHs. For smaller values of $G \mu$ we would require some amount of Eddington accretion in order to explain the number density of SMBHs. In that case, the predicted number of IMBH candidates would be smaller than what is given in (\ref{IMBHnumber}) unless the smaller loops also undergo similar Eddington accretion.


We have normalized our calculations to yield one SMBH of mass of at least $10^6 \Msol$ per galaxy. As long as the loops are created in the radiation phase, the number density $n(>M)$ of  supermassive seed scales with the seed mass greater than $M$ (using the fact that the linear accretion factor for all of these loops is the same and hence $M$ is proportional to the loop radius $R$) as\footnote{Note that $n(M)$ is the number density of black holes per unit mass, and $n_{>M}$ is the number density of SMBH seeds for masses greater than $10^6 \, M_{\odot}$.}
\begin{align}
n_{>M} \, \propto \, M^{-3/2} \, .
\end{align}

For loops created in the matter period, we have (see (\ref{matterdist})) $n(R) \sim R^{-2}$. Furthermore, the linear growth factor is reduced since formation after matter-radiation equality means less time to accrete. In particular, the linear growth factor, $GF$, now depends on $R$
\begin{align}
GF(R) \, = \, GF(R_c) (z_{eq}+1) \left(\frac{t_{eq}}{R} \right)^{2/3} \, ,
\end{align}
where $R_c$ is the radius of the loop formed at the time $t_{eq}$. Hence, taking into account this radius-dependent growth factor we have
\begin{align}
n(M) \, \propto \, M^{-4} \, ,
\end{align}
and hence the number density of seeds with mass greater or equal to $M$ scales as
\begin{align}
n_{>M} \, \propto \, M^{-3}
\end{align}
Thus, the mean separation $d_M$ of seeds with mass greater or equal to $M$ (for large masses) scales as
\begin{align}
d_M \, = d_6 \frac{M}{M_6} \left( \frac{M_6}{M_{eq}} \right)^{1/2} \, , \label{distscaling}
\end{align}
where $M_{eq}$ is the mass of a seed from a loop created at time $t_{eq}$, and we use $M_6 = 10^6 \Msol$ and $d_6$ to be the mean separation of the corresponding seeds.


Making use of $R_c = t_{eq}$ (loops form with size of order the Hubble radius) and the value of $G\mu$ from (\ref{muvalue}) we obtain 
\begin{align}
M_{eq} \, \simeq \, \beta \mu t_{eq} (z_{eq} + 1) {\cal{E}}_{eq} \,
\sim \, 2.4 \times 10^8 N^{-2/3} \frac{{\cal{E}}_{eq}}{{\cal{E}}_1} \Msol \, ,
\end{align}
where ${\cal{E}}_{eq}$ is the Eddington growth rate of loops formed at $t_{eq}$. Hence, from (\ref{distscaling}) we get
\begin{align}
d_9  \, \sim \, 0.3 {\rm{Gpc}} N^{1/3} \left( \frac{{\cal{E}}_1}{{\cal{E}}_{eq}} \right)^{1/2}
\end{align}
for the mean separation of seeds which can accrete SMBHs of mass $10^9 \Msol$. 

Assuming no Eddington accretion or an Eddington accretion factor independent of mass for these superheavy objects, and using the value $N = 6 \times 10^{-3}$ from numerical simulations \cite{NGsimuls}, our model thus predicts one SMBH of mass greater than $10^9 \Msol$ per volume $d_9^3$ with $d_9 \simeq  60 {\rm{Mpc}}$, which agrees well with the observed separation of such black hole monsters \cite{Marta2}.

\section{Conclusions and Discussion} \label{conclusion}

We have studied the implications of the proposal that both super-massive black holes (SMBHs) and intermediate black holes (IMBHs) could originate from cosmic string loop seeds. The cosmic string model contains (in principle) one free parameter, namely the string tension. Normalizing the string tension to yield one candidate seed per galaxy which can develop into a SMBH of mass greater or equal to $10^6 \Msol$, we predicted the number per galaxy of IMBHs capable of seeding black holes in the ``mass gap'' window of $65 \, M_{\odot} - 130 \, M_{\odot}$. This number is $10^6$ modulo Eddington accretion factors. We can also predict the mean separation $d_9$ of loops capable of seeding monster SMBHs of mass greater than $10^9 \Msol$. We obtain $d_9 \sim 60 {\rm Mpc}$.

Our model predicts a continuum of black hole masses inside any galaxy, from one SMBH with mass greater or equal to $10^6 \Msol$ down to a lower cutoff mass $M_c$ given by
\begin{align}
M_c \, \sim \, \gamma G\mu \mu t_{eq} (z_{eq}+1) 
\end{align}
(modulo Eddington accretion factors). Would-be black holes with a smaller mass than this would have to be formed from string loops that would have decayed by matter-radiation equality, and therefore would not have undergo any efficient accretion. Inserting the value of the string tension from (\ref{muvalue}), we find
\begin{align}
M_c \, \sim \, 10^{-2} \, \gamma (z_{eq}+1)^{-1/2} {\cal{E}}_1^{-2} \Msol \, ,
\end{align}
which, using the value of $\gamma \sim 10^2$ from studies of gravitational radiation from string loops \cite{VV}, is about $10^{-2}\Msol$, assuming no Eddington accretion.

We expect that the accretion onto these cosmic string loops collapses into a black hole before structure formation takes place. Therefore, the formation and merger history of binary systems of these IMBHs will closely resemble that of primordial black holes (see for example \cite{PBHmerger})

Note that in our scenario, the string loop-seeded black holes provide a negligible fractional contribution $\Omega_{BH}$ to the dark matter density 
\begin{align}
\Omega_{BH} \simeq \, 12 \pi \times \beta \gamma^{-1} (G\mu)^{1/2} N\, \sim \, 0.4 \times 10^{-2} N^{2/3}\, ,
\end{align}
being dominated by black holes of mass near the cutoff mass $M_c$, and hence our model is consistent with observational bounds on the black hole contribution to the total energy budget of the universe (see e.g. \cite{Bertone}).

\section*{Acknowledgement}

\noindent RB wishes to thank the Pauli Center and the Institutes of Theoretical Physics and of Particle- and Astrophysics of the ETH for hospitality. The research of RB and HJ at McGill is supported in part by funds from NSERC and from the Canada Research Chair program. BC thanks support from a Vanier-CGS fellowship and from NSERC.  RB wishes to thank Ken Olum and Marta Volonteri for discussions. JH acknowledges fellowships from Hydro-Quebec (through the McGill Physics Department) and from the McGill Space Institute.

\end{document}